\documentclass[cits]{PoS}
\usepackage{axodraw}
\usepackage{graphicx}
\usepackage{epsfig}
\usepackage{psfrag}
\usepackage{soul}
\usepackage{multirow}
\usepackage{mathptmx}
\usepackage{mathrsfs}
\usepackage{amsmath, amssymb}
\newcommand{\xiav}{\langle\,\xi\,\rangle}
\newcommand{\xisqav}{\langle\,\xi^2\,\rangle}

\title{Lattice Results for Vector Meson Couplings and
Parton Distribution Amplitudes}

\ShortTitle{Parton Distribution Amplitudes}

\author{UKQCD and RBC Collaborations}

\author{M.A. Donnellan, J. Flynn, A. J\"uttner,
\speaker{C.T. Sachrajda}\\
UKQCD Collaboration
School of Physics and Astronomy, University of Southampton,
Southampton
SO17 1BJ UK\\
E-mail: \email{mad2@phys.soton.ac.uk},
\email{jflynn@phys.soton.ac.uk},
\email{juettner@phys.soton.ac.uk}, \email{cts@phys.soton.ac.uk}}

\author{D. Antonio, P.A. Boyle, C. Maynard, B. Pendleton, R.Tweedie\\
School of Physics, The University of Edinburgh,
Edinburgh, EH9 3JZ UK \\
        E-mail: \email{s0459477@sms.ed.ac.uk}, \email{paboyle@ph.ed.ac.uk}
        \email{c.maynard@ed.ac.uk}, \email{bjp@ph.ed.ac.uk}, \email{rjt@ph.ed.ac.uk}}

\abstract{We present results for the couplings of light vector
mesons to vector and tensor currents and on the low moments of the
light-cone distribution amplitudes of the pion and kaon. The
calculations are performed on the RBC and UKQCD collaborations'
ensembles generated with the Iwasaki gauge action and with 2+1
flavours of domain wall fermions. The (preliminary) results for
the ratios of the couplings of the vector meson to the vector and
tensor currents ($f_V$ and $f^T_V$ respectively) in the
$\overline{\textrm{MS}}$ scheme at 2\,GeV are:
$f_\rho^T/f_\rho=0.681(20)$; $f_{K^\ast}^T/f_{K^\ast}=0.712(11)$
and $f_\phi^T/f_\phi=0.751(9)$. For the first moment of the kaon's
distribution amplitude we find (in the same scheme and at the same
scale) $\xiav_K=0.029(2)$ and for the second moment
$\xisqav_\pi=0.28(3)$ and $\xisqav_K=0.27(2)$\,.}

\FullConference{The XXV International Symposium on Lattice Field Theory\\
         July 30-4 August 2007\\
         Regensburg, Germany}

\begin{document}

\section{Introduction}\label{sec:intro}

In this talk I briefly review the status of calculations by the
RBC and UKQCD collaborations of the couplings of light vector
mesons to vector and tensor currents and of the low moments of the
light-cone distribution amplitudes of the pion and kaon. (A
detailed description will be presented in forthcoming
publications.) We use the collaborations' two principal ensembles,
generated with the Iwasaki gauge action and domain wall fermions
with an inverse lattice spacing of about 1.73\,GeV. The two
lattices have $24^3\times 64\times 16$ points (corresponding to a
physical volume of $L\simeq 2.7\,$fm) and $16^3\times 32\times 16$
points (physical volume, $L\simeq 1.8\,$fm). The properties of the
$16^3$ (spatial) lattice have been presented in detail in
ref.\,\cite{sixteencubed} and those of the $24^3$ one have been
summarised in the plenary talk by Peter Boyle at this
conference~\cite{pab} and the corresponding paper is being
prepared for publication.

On the $24^3$ lattice, measurements have been made using 4 values
of the light-quark mass whose bare values (and the corresponding
pion masses) are as follows:
\begin{eqnarray*}
ma=0.03\ (m_\pi\simeq 670\,\textrm{MeV}); &\quad&
ma=0.02\ \ \,(m_\pi\simeq 555\,\textrm{MeV});\\
ma=0.01\ (m_\pi\simeq 415\,\textrm{MeV}); &\quad& ma=0.005\
(m_\pi\simeq 330\,\textrm{MeV})\,,\end{eqnarray*} where $a$ is the
lattice spacing. On the $16^3$ lattice results were obtained with
$ma=0.03,\,0.02$ and $0.01$. For the strange quark we take
$m_sa=0.04$ and the residual mass is
$m_{\textrm{res}}a=0.00325(2)$\,.

On the $16^3$ lattice it was found that $a^{-1}=1.62(4)\,$GeV and
$m_sa=0.04$~\cite{sixteencubed}. With the new data on the $24^3$
lattice it is possible to reach lower masses and perform a more
detailed investigation of the chiral behaviour~\cite{pab,enno}.
The conclusion is that $a^{-1}=1.73(3)\,$GeV and $m_sa\simeq
0.0344(16)$ and below I will discuss the corresponding corrections
to the results obtained directly with $m_sa=0.04$.

\section{Couplings of Vector Mesons to Vector and Tensor Currents}
\label{sec:frho}

The couplings of the vector meson to the vector and tensor
currents ($f_V$ and $f_V^T$ respectively) are defined through the
matrix elements:
\begin{eqnarray}
\langle\,0\,|\,\bar{q}_2(0)\gamma^\mu
q_1(0)\,|\,V(p;\lambda)\,\rangle
&=&f_V\,m_V\,\varepsilon_\lambda^\mu\\
\langle\,0\,|\,\bar{q}_2(0)\sigma^{\mu\nu}
q_1(0)\,|\,V(p;\lambda)\,\rangle
&=&if_V^T(\mu)\,\left(\varepsilon_\lambda^\mu p^\nu-
\varepsilon_\lambda^\nu p^\mu\right)\,,
\end{eqnarray}
where $p$ and $\lambda$ are the momentum and polarization state of
the vector meson $V(p;\lambda)$ and $\varepsilon_\lambda$ is the
corresponding polarization vector. The tensor bilinear operator
$\bar{q}_2\sigma^{\mu\nu} q_1$ (and hence $f_V^T(\mu)$) depends on
the renormalization scheme and scale $\mu$. The final results will
be quoted in the $\overline{\textrm{MS}}$ scheme at $\mu=2$\,GeV.

\subsection{Experimental Determination of $f_V$}

The decay constants $f_V$ can be determined experimentally. For
the charged $\rho$ and $K^\ast$ mesons, one can use $\tau$ decays
to deduce $f_{\rho}$ and $f_{K^\ast}$ as illustrated by the
following diagram:

\begin{center}\begin{picture}(100,40)(-5,0)
\ArrowLine(0,30)(30,30)\ArrowLine(30,30)(60,30)
\ArrowLine(50,20)(70,10)\SetColor{Magenta}\Photon(30,30)(50,20){1.5}{5}
\GCirc(30,30){1.5}{0}\GCirc(50,20){1.5}{0}
\Text(74,10)[l]{\small$\rho,\,K^\ast$}
\Text(64,30)[l]{\small$\nu_\tau$} \Text(-4,30)[r]{\small$\tau$}
\end{picture}\end{center}
From the measured branching ratios one obtains the following
values for the decay constants~\cite{blmt}:
\begin{eqnarray}
\textrm{Br}(\tau^-\to\rho^-\nu_\tau)=(25.0\pm
0.3)\%&\Rightarrow&f_{\rho^-}\,\simeq 208\,\textrm{MeV}\\
\textrm{Br}(\tau^-\to K^{\ast\,-}\nu_\tau)=(1.29\pm
0.03)\%&\Rightarrow&f_{K^{\ast\,-}}\!\!\simeq 217\,\textrm{MeV}
\end{eqnarray}
One can also determine $f_{\rho^0}$ from the width of the decay of
the $\rho^0$ into $e^+e^-$ which gives $f_{\rho^0}=216(5)$\,MeV.
Similarly from the width of the decay $\phi\to e^+e^-$ one deduces
$f_\phi\simeq 233$\,MeV\,.

The couplings $f_V^T$ are not known from experiment but are used
as inputs in sum-rule calculations and other phenomenological
applications. I will present our results for $f_V^T/f_V$, which
can then be combined with the experimental values of $f_V$ to
obtain $f_V^T$. For the $\phi$ we neglect the Zweig suppressed
disconnected contribution.

\subsection{Lattice Calculation of
$f_V^T/f_V$}\label{subsec:calculation_fv}

In order to determine $f_V^T/f_V$ it is sufficient to calculate
the following zero-momentum correlation functions:
\begin{eqnarray}
C_{V^{s_1}V^{s_2}}(t)&\equiv&\sum_{\vec{x},i}
\langle\,0\,|V_i^{s_1}(t,\vec{x})\,V_i^{s_2}(0)\,|\,0\,\rangle =
3f_V^{s_1}
f_V^{s_2}m_Ve^{-m_VT/2}\cosh\left(m_V\left(T/2-t\right)\right)\\
C_{V^{s_1}T^{s_2}}(t)&\equiv&\sum_{\vec{x},i}
\langle\,0\,|V_i^{s_1}(t,\vec{x})\,T_{4i}^{s_2}(0)\,|\,0\,\rangle
=3f_V^{s_1}
f_V^{Ts_2}m_Ve^{-m_VT/2}\sinh\left(m_V\left(T/2-t\right)\right)\,,
\end{eqnarray}
where $V_\mu$ and $T_{\mu\nu}$ represent the vector and tensor
current and $s_1$ and $s_2$ label the smearing at the sink and
source respectively. $i$ is a spatial index ($i=1,2,3$). From the
ratio
\begin{equation}R(t)=
\frac{C_{V^{s_1}T^{L}}}{C_{V^{s_1}V^{L}}}=\frac{f_V^T}{f_V}\,
\tanh(m_V(T/2-t))\end{equation} we readily obtain the ratio of
(bare) couplings. In practice we also calculate the correlation
functions $C_{T^{s_1}V^{s_2}}(t)$ and include them in the
analysis. For the smearing we use a gauge invariant Gaussian and
hydrogen-like gauge-fixed wave function.

\subsection{Results}\label{subsec:fvresults}

\begin{table}
\begin{center}
\begin{tabular}{lcccccc}
\hline\hline\\ [-4mm]
&Volume&$ma=0.03$&$ma=0.02$&$ma=0.01$&$ma=0.005$\\
\hline\\ [-4mm]
$f_\rho^T/f_\rho$&$24^3$&0.6813(47)&0.6699(76)&0.648(12)&0.624(23)\\
[1mm]
$f_\rho^T/f_\rho$&$16^3$&0.6885(63)&0.6660(69)&0.6236(80)&-\\
\hline\\ [-4mm]
$f_{K^\ast}^T/f_{K^\ast}$&$24^3$&0.6872(70)&0.6826(61)&0.6673(43)&0.6556(60)\\
[1mm]$f_{K^\ast}^T/f_{K^\ast}$&$16^3$&0.6879(48)&0.6774(45)&0.6516(46)&-\\
\hline\\ [-4mm]
$f_\phi^T/f_\phi$&$24^3$&0.6929(49)&0.6942(40)&0.6876(24)&0.6868(28)\\
[1mm]$f_\phi^T/f_\phi$&$16^3$&0.7023(47)&0.6907(31)&0.6822(33)&-\\
\hline\hline\\ %[-4mm]
\end{tabular}
\caption{Results for the bare ratios of couplings $f_V^T/f_V$
obtained on both the lattice volumes.\label{tab:frho}}
\end{center}\end{table}

In table~\ref{tab:frho} we present the bare values of $f_V^T/f_V$
corresponding to both the lattice volumes. It can be seen that the
measured results are obtained with excellent precision. For the
two heaviest quark masses the results from the two volumes are
fully compatible; for $ma=0.01$ perhaps there is a hint of a small
finite volume effect and we do not include the results from the
$16^3$ lattice at this mass in the final analysis.

\begin{figure}\begin{center}
 \psfrag{xlabel}[t][b][0.8][0]{$(m+m_{\textrm{res}})a$}
 \psfrag{ylabel}[b][b][0.8][0]{$f_\rho^T/f_\rho$}
 \epsfig{scale=.17,angle=-90,file=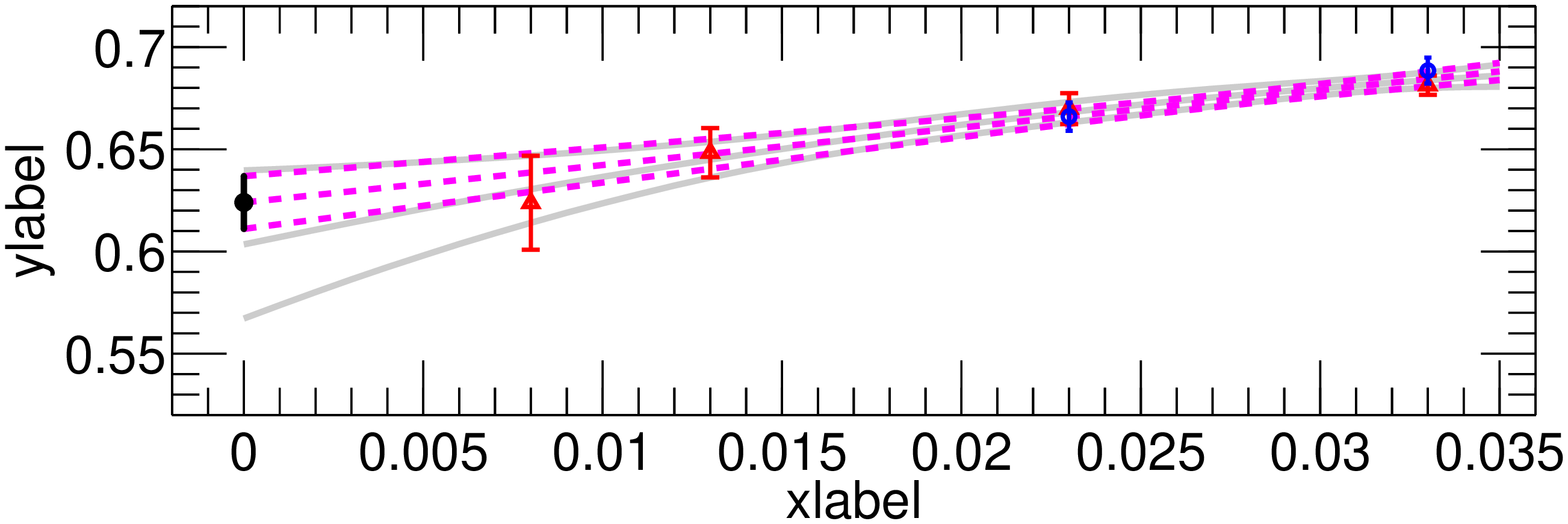}
\quad \psfrag{ylabel}[b][c][0.75][0]{$f_{K^\ast}^T/f_{K^\ast}$}
 \epsfig{scale=.17,angle=-90,file=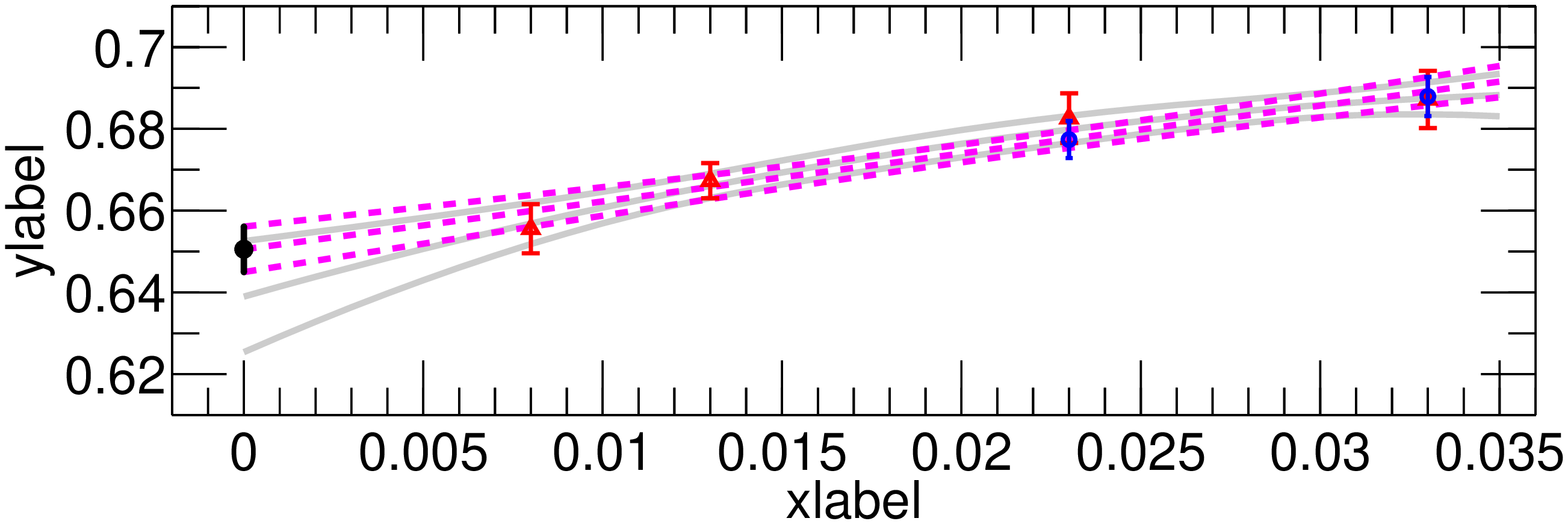}
\quad \psfrag{ylabel}[b][b][0.75][0]{$f_\phi^T/f_\phi$}
 \epsfig{scale=.17,angle=-90,file=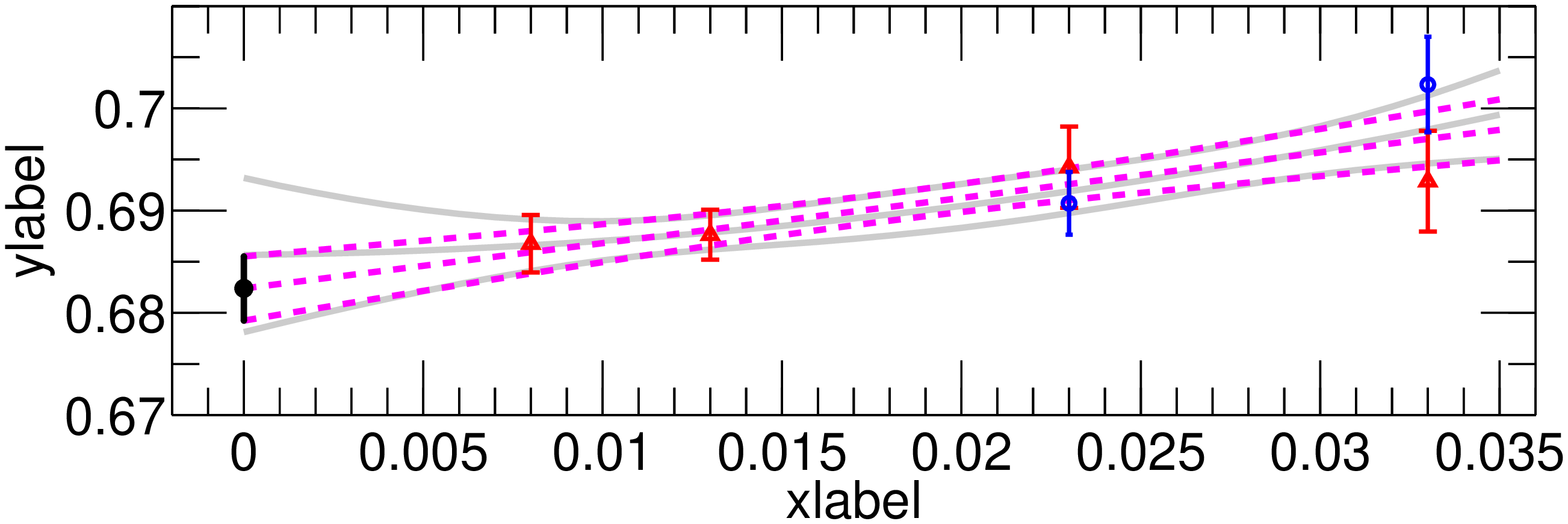}
\end{center}\caption{Chiral extrapolations for $f_\rho^T/f_\rho$, $f_{K^\ast}^T/f_{K^\ast}$
and $f_\phi^T/f_\phi$ respectively. The broken red lines represent
a linear fit to the mass behaviour and the solid grey lines a
quadratic fit.\label{fig:chiral}}\end{figure}

From fig.~\ref{fig:chiral} it can be seen that the dependence of
the bare $f_V^T/f_V$ on the masses of the light quarks is very
mild.
 For the ratio of
bare couplings in the chiral limit we obtain:
\begin{equation}\frac{f_\rho^T}{f_\rho}=0.624(13)_{(-21)}^{(+0)};\quad
\frac{f_{K^\ast}^T}{f_{K^\ast}}=0.6505(56)_{(-115)}^{(+0)};\quad
\frac{f_{\phi}^T}{f_\phi}=0.6824(31)^{(+35)}_{(-0)}\,,\label{eq:barefv}
\end{equation}
where the central value corresponds to the linear extrapolation
and the second error is the difference between the results from
the linear and quadratic extrapolations.

The bare results in eq.\,(\ref{eq:barefv}) were obtained with the
``notional" strange quark mass of $m_sa=0.04$ rather than the
updated value of 0.0344\,. The values of the ratios in
eq.\,(\ref{eq:barefv}) are very similar for the $\rho,\ K^\ast$
and $\phi$ mesons and we correct for the change in $m_s a$ by
linear extrapolation in the valence quark mass
($am_s^{\textrm{sea}}$ is fixed at 0.04). Thus, for example, for
the $K^\ast$ meson we write:
\begin{equation}
\frac{f_{K^\ast}^T}{f_{K^\ast}}(m_sa=0.0344)=
\frac{f_{K^\ast}^T}{f_{K^\ast}}(m_sa=0.04)+\frac{\Delta}{(0.04+am_{\textrm{res}})}\,
(0.0344-0.04)\,,
\end{equation}
where $\Delta=
f_{K^\ast}^T/f_{K^\ast}(m_sa=0.04)-f_\rho^T/f_\rho$. The corrected
bare values are then
\begin{equation}\frac{f_\rho^T}{f_\rho}=0.624(13)_{(-21)}^{(+0)};\quad
\frac{f_{K^\ast}^T}{f_{K^\ast}}=0.6470(56)_{(-115)}^{(+0)};\quad
\frac{f_{\phi}^T}{f_\phi}=0.6748(31)^{(+35)}_{(-0)}(30)\,,\label{eq:barefvcorr}
\end{equation}
where we have added the third error on $f_{\phi}^T/f_\phi$ to
reflect the uncertainty on the strange quark mass and to allow for
non-linearities in the behaviour (for the $K^\ast$ the
corresponding error is negligible).

We determine the renormalization constants non-perturbatively
using the Rome-Southampton method and run the results to
2\,GeV~\cite{nprpaper}:
\begin{equation}
\frac{f_V^T(2\,\textrm{GeV})}{f_V}=\frac{Z_T(2\,\textrm{GeV}a)}{Z_V}
\,\frac{f_V^{T\,\textrm{bare}}(a)}{f^{\textrm{bare}}_V}=1.11(1)\,
\frac{f_V^{T\,\textrm{bare}}(a)}{f^{\textrm{bare}}_V}\,.\end{equation}
In the $\overline{\textrm{MS}}$ scheme with $\mu=2$\,GeV we
finally obtain:
\begin{equation}\frac{f_\rho^T}{f_\rho}=0.681(20);\quad
\frac{f_{K^\ast}^T}{f_{K^\ast}}=0.712(11);\quad
\frac{f_{\phi}^T}{f_\phi}=0.751(9)\,.\label{eq:finalpreliminary}
\end{equation}

These results can be compared with previous quenched lattice
results:

\begin{center}\begin{tabular}{c|c|c|c}
Reference&$\frac{f_\rho^T(2\,\textrm{GeV})}{f_\rho}$&
$\frac{f_{K^\ast}^T(2\,\textrm{GeV})}{f_{K^\ast}}$&
$\frac{f_\phi^T(2\,\textrm{GeV})}{f_\phi}$\\
\hline Becirevic et
al.~\cite{blmt}&$0.720\pm0.024^{+0.016}_{-0.000}$&
$0.739\pm 0.017^{+0.003}_{-0.000}$&$0.759\pm 0.009\pm0.000$\\
Braun et al.~\cite{braun}&$0.742\pm0.014$&$-$&$0.780\pm0.008$\\
\hline
\end{tabular}\end{center}

The QCDSF/UKQCD also presented the result $f_\rho^T=168(3)$\,MeV
at Lattice 2005, using an $N_f=2$ $O(a)$ improved clover action
with a range of lattice spacings
($0.07<a<0.11$\,fm)~\cite{ukqcdqcdsf}. Combining our result for
the ratio from eq.\,(\ref{eq:finalpreliminary}) together with the
experimental value for $f_\rho$ we obtain a smaller value
$f_\rho^T=140(5)$\,MeV\,.

\section{Pion and Kaon Distribution Amplitudes}
\label{sec:pda}

The leading-twist distribution amplitude of the pion, $\phi_\pi$,
is defined by:
\begin{equation}
\left.\langle\,\pi^+(q)\,|\,\bar{u}_\alpha(z)\,{\cal
P}(z,-z)\,d_\beta(-z)\,|\,0\,\rangle
\right|_{z^2=0}\equiv\frac{if_\pi}{4}\,(\not\!{q}\gamma_5)_{\beta\alpha}\,
\int_0^1du\,e^{i(2u-1)q\cdot z}\phi_\pi(u,\mu)\,,\end{equation}
where ${\cal P}$ is the path-ordered exponential. For the kaon
there is a similar definition with the obvious change of quark
flavours. The (universal) distribution amplitudes contain the
non-perturbative QCD effects in hard exclusive processes.

Performing a series expansion to first order in $z$:
\begin{equation}\label{eq:xidef}
\langle\,\pi^+(q)\,|\,\bar{u}(0)\,\gamma_5\gamma_{\{\rho}\overset{\leftrightarrow}{D}_{\mu\}}
\,d(0)\,|\,0\,\rangle =f_\pi\,(iq_\mu)(iq_\rho)
\int_0^1du\,(2u-1)\,\phi_\pi(u,\mu)\equiv
f_\pi\,(iq_\mu)(iq_\rho)\xiav_\pi\,.\end{equation} $\xiav_\pi=0$
by isospin symmetry; we calculate $\xiav_K$.

Performing the series expansion to next order we have:
\begin{equation}
\langle\,\pi^+(q)\,|\,\bar{u}\,\gamma_5\gamma_{\{\rho}\,\overset{\leftrightarrow}{D}_{\mu}
\overset{\leftrightarrow}{D}_{\nu\}}\,d\,|\,0\,\rangle
=f_\pi\,(iq_\rho)(iq_\mu)(iq_\nu)
\int_0^1du\,(2u-1)^2\,\phi_\pi(u,\mu)\,.\end{equation} We define
$\xisqav_\pi\equiv\int_0^1du\,(2u-1)^2\,\phi_\pi(u,\mu)$ and
similarly for the kaon.

The lattice determination of $\xiav^{\textrm{bare}}_K$ is very
straightforward. We compute the ratio of two point correlation
functions:
\begin{equation}
\frac{\sum\limits_{\vec x}e^{i \vec{p}\cdot\vec{x}}
  \langle 0|O_{\{\rho\mu\}}(t,\vec{x}\,) P^\dagger(0)|0\rangle}
{\sum\limits_{\vec x}e^{i\vec{p}\cdot\vec{x}}
  \langle 0|A_\nu(t,\vec{x}\,) P^\dagger(0)|0\rangle}\quad
  \xrightarrow[t\ \textrm{large}]{}\quad
  i\,\frac{p_\rho p_\mu}{p_\nu}\ \xiav^{\textrm{bare}}_K\ ,
\end{equation} where $P$ and $A_\nu$ are the pseudoscalar
and axial bilinears and
$O_{\{\rho\mu\}}=\bar{u}\,\gamma_5\gamma_{\{\rho}
\overset{\leftrightarrow}{D}_{\mu\}} \,s$. The most appropriate
choice of indices is $O_{\{4i\}}$ and $\nu=4\,.$ The second moment
is evaluated from a similar ratio with
$O_{\{4ij\}}=\bar{u}\gamma_5\gamma_{\{4}\,
\overset{\leftrightarrow}{D}_{i}
\overset{\leftrightarrow}{D}_{j\}}s$ and $i\neq j$\,. The
transformation properties of the operators under the lattice
symmetries have been studied in
detail~\cite{ms87,qcdsfgrouptheory}\,.

\subsection{Results for
$\xiav_K$}\label{subsec:xiav}

\begin{figure}\begin{center}
\psfrag{ms-minus-mq}[t][c][1][0]{\footnotesize$am_s-am_q$}
  \psfrag{fstmom}[b][t][1][0]{\footnotesize$\xiav_K^{\rm bare}$}
  \psfrag{Legendmass4}[c][c][0.8][0]{\footnotesize$am_{ud}=0.005$}
  \psfrag{Legendmass1}[c][c][0.8][0]{\footnotesize$am_{ud}=0.01$}
  \psfrag{Legendmass2}[c][c][0.8][0]{\footnotesize$am_{ud}=0.02$}
  \psfrag{Legendmass3}[c][c][0.8][0]{\footnotesize$am_{ud}=0.03$}
  \epsfig{scale=.25,angle=270,file=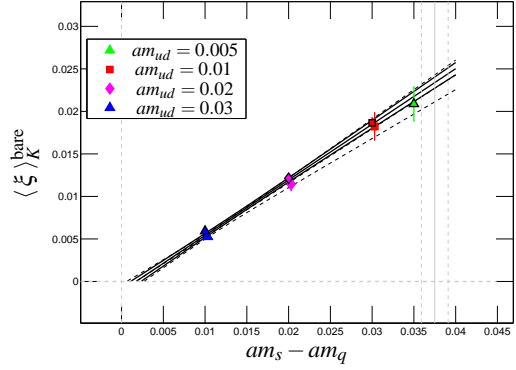}
\caption{Bare values of $\xiav_K$ vs the quark mass. The physical
region $m_sa-m_qa=0.0375(16)$ is marked.
\label{fig:xiavbare}}\end{center}\vspace{-0.2in}\end{figure}

In fig.\,\ref{fig:xiavbare} we plot our bare results as a function
of the mass of the light quark with $m_sa=0.04$. The three points
without a black border were obtained on the $16^3$ lattice and the
results were presented in ref.\,\cite{xiav2006}. The remaining
four points (including the point at $ma=0.005$) were obtained from
the $24^3$ lattice and there is no evidence of significant finite
volume effects. $\xiav_K$ is clearly non-zero away from the
$SU(3)$ limit and the results are compatible with the prediction
from lowest order chiral perturbation theory that $\xiav_K$ is
proportional to $m_s-m_q$ without logarithms~\cite{chenstewart}\,.
The linearly extrapolated value in the flavour symmetry limit
($ma=m_sa=0.04$) is approximately 0. The bare result in the
physical limit ($(m_s-m_q)a=0.0375(16)$, see
fig.\,\ref{fig:xiavbare}) obtained with a linear fit is
$\xiav_K^{\rm bare}=0.0234(12)$. Multiplying this result by the
renormalization factor (which, so far we have only calculated
perturbatively) we finally obtain:
\begin{equation}
\xiav_K^{\overline{\textrm{MS}}}(2\,\textrm{GeV}) = 0.029(2)\,\,.
\end{equation}
\subsection{Results for $\xisqav$}
\begin{figure}
\begin{minipage}{.48\linewidth}\begin{center}
\psfrag{mqplusmres}[t][c][1][0]{\footnotesize$am_q+am_{\rm res}$}
  \psfrag{scndmom}[b][c][1][0]{\footnotesize$\xisqav_\pi^{\rm bare}$}
  \psfrag{Legendmass1}[c][c][0.8][0]{\footnotesize$am_{ud}=0.01$}
  \psfrag{Legendmass2}[c][c][0.8][0]{\footnotesize$am_{ud}=0.02$}
  \psfrag{Legendmass3}[c][c][0.8][0]{\footnotesize$am_{ud}=0.03$}
  \psfrag{Legendmass4}[c][c][0.8][0]{\footnotesize$am_{ud}=0.005$}
  \epsfig{scale=.25,angle=270,file=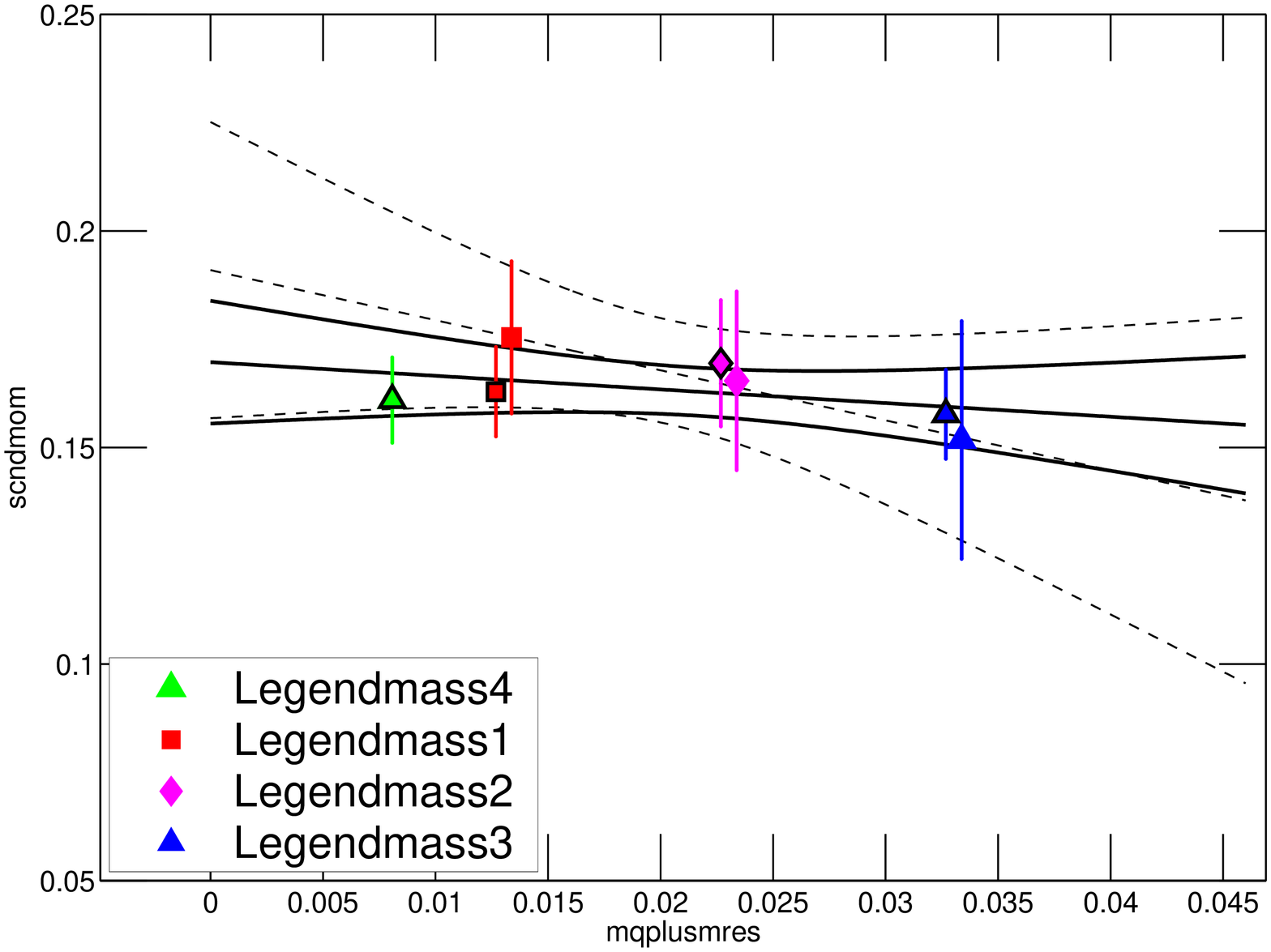}
\end{center}\end{minipage}\hspace{0.039\linewidth}
\begin{minipage}{.48\linewidth}\begin{center}
  \psfrag{mqplusmres}[t][c][1][0]{\footnotesize$am_q+am_{\rm res}$}
  \psfrag{scndmom}[b][c][1][0]{\footnotesize$\xisqav_K^{\rm bare}$}
  \psfrag{Legendmass1}[c][c][0.8][0]{\footnotesize$am_{ud}=0.01$}
  \psfrag{Legendmass2}[c][c][0.8][0]{\footnotesize$am_{ud}=0.02$}
  \psfrag{Legendmass3}[c][c][0.8][0]{\footnotesize$am_{ud}=0.03$}
  \psfrag{Legendmass4}[c][c][0.8][0]{\footnotesize$am_{ud}=0.005$}
  \epsfig{scale=.25,angle=270,file=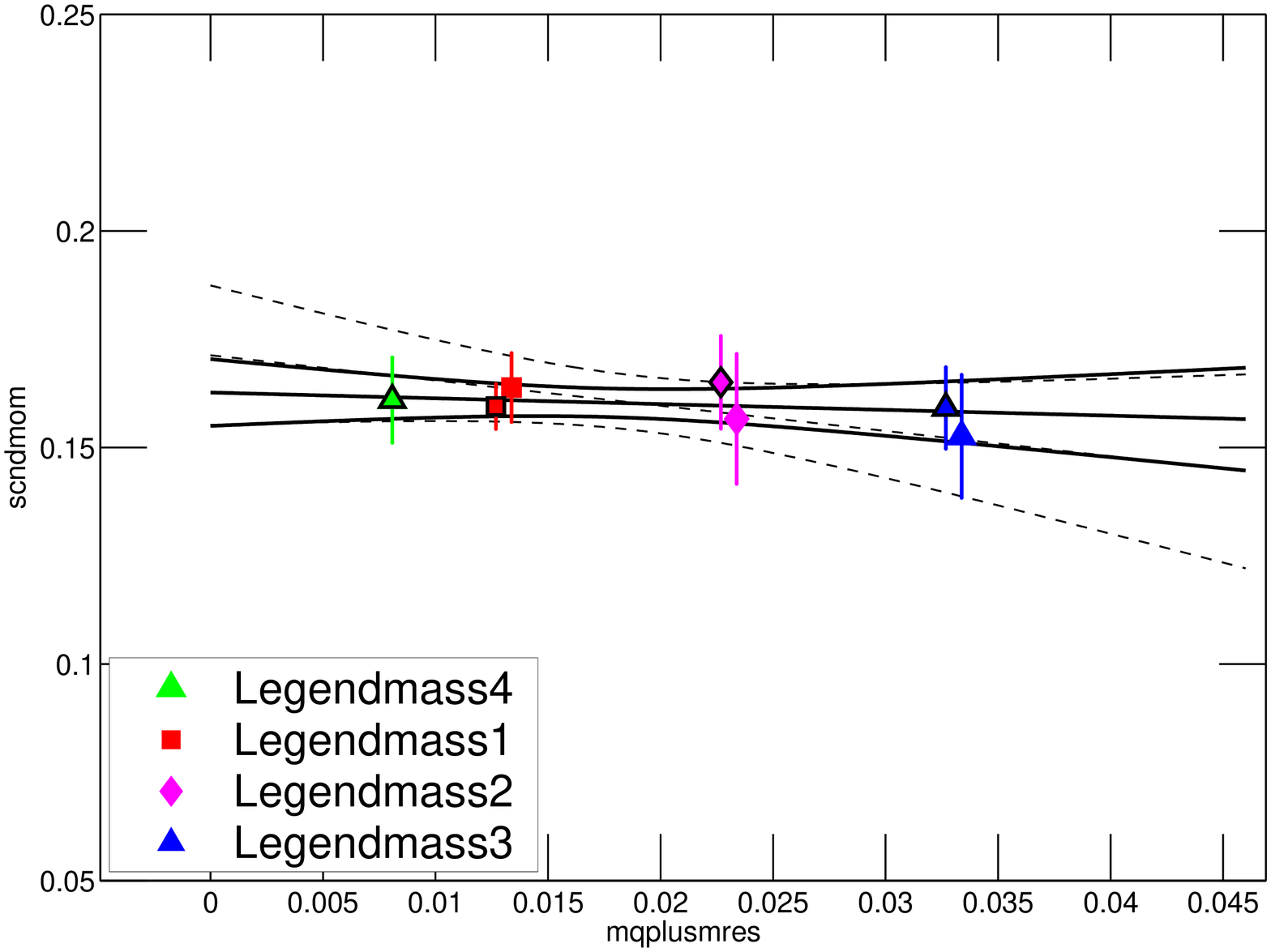}
\end{center}\end{minipage}\caption{Bare values of $\xisqav_\pi$ and
$\xisqav_K$ as a function of the quark
mass.\label{fig:xisqav}}\end{figure} Figure\,\ref{fig:xisqav}
contains our results for $\xisqav^{\textrm{bare}}_\pi$ and
$\xisqav^{\textrm{bare}}_K$ from which we deduce the values in the
chiral limit: $\xisqav^{\textrm{bare}}_\pi=0.171(15)$ and
$\xisqav^{\textrm{bare}}_K=0.163(8)$\,. In order to obtain the
results in the $\overline{\textrm{MS}}$ scheme, we need to perform
the renormalization, which up to now we have done perturbatively.
The operator
$O_{DD}=\bar\psi\gamma_{\{\mu}\gamma_5\overset{\leftrightarrow}D_\nu
\overset{\leftrightarrow}D_{\kappa\}\psi}$ (with $\mu, \nu$ and
$\kappa$ all distinct) mixes with $O_{\partial\partial}=
\partial_{\{\nu}\partial_\kappa\,\bar\psi
\gamma_{\mu\}}\gamma_5\psi$. Dividing by the matrix element of the
axial current we find
\begin{equation}
\xisqav^{\overline{\textrm{MS}}}(a)=1.52(7)\xisqav^{\textrm{bare}}
+0.022(7)\,.
\end{equation}
Inserting the measured values of $\xisqav^{\textrm{bare}}$ and
(two-loop) running the result to 2\,GeV we obtain
\begin{equation}
\xisqav^{\overline{\textrm{MS}}}_\pi(2\,\textrm{GeV})=0.278\pm0.026
\quad\textrm{and}\quad
\xisqav^{\overline{\textrm{MS}}}_K(2\,\textrm{GeV})=0.267\pm
0.018\,.
\end{equation} Since $\xisqav_\pi=\xisqav_K$ to excellent
precision, we do not make any interpolation to $m_sa=0.0344$.
\subsection{Final Comments}
We have shown that lattice calculations of $\xiav_K,$ and
$\xisqav_{\pi,K}$ can be performed with an excellent precision.
Our (preliminary) results are:
\begin{equation}
\xiav_K^{\overline{\textrm{MS}}}(2\,\textrm{GeV})=0.029(2)\,,\quad
\xisqav^{\overline{\textrm{MS}}}_\pi(2\,\textrm{GeV})=0.28(3)\,,\quad
\xisqav^{\overline{\textrm{MS}}}_K(2\,\textrm{GeV})=0.27(2)\,.
\end{equation}
Important improvements will be to perform the non-perturbative
renormalization and to repeat the calculation at a finer lattice
spacing (we estimate a further 5\% error due to discretization
effects). Our results are in agreement with those of an $N_f=2$
study using Improved Wilson fermions~\cite{qcdsfpda}
\begin{equation}
\xiav_K^{\overline{\textrm{MS}}}(2\,\textrm{GeV})=0.0272(5)\,,\quad
\xisqav^{\overline{\textrm{MS}}}_\pi(2\,\textrm{GeV})=0.269(39)\,,\quad
\xisqav^{\overline{\textrm{MS}}}_K(2\,\textrm{GeV})=0.260(6)\,.\end{equation}

The analysis of moments of the distribution amplitudes of vector
mesons is in progress.


\begin{thebibliography}{99}
\bibitem{sixteencubed} C.~Allton {\it et al.}  [RBC \& UKQCD Collaborations],
\textit{2+1 flavor domain wall QCD on a (2-fm)$^3$ lattice: Light
meson spectroscopy with $L_s = 16$},
  Phys.\ Rev.\  D {\bf 76} (2007) 014504
   \texttt{[arXiv:hep-lat/0701013]}.
\bibitem{pab} P.A.~Boyle, \textit{2+1 flavour domain wall fermion simulations
by the RBC and UKQCD collaborations}, these proceedings.
\bibitem{enno} M.~Lin and E.E.~Scholz, \textit{Chiral limit and
light quark masses in 2+1 flavor domain wall QCD}, these
proceedings.
\bibitem{blmt} D.~Becirevic, V.~Lubicz, F.~Mescia and C.~Tarantino,
  \textit{Coupling of the light vector meson to the vector and to the tensor
  current,}
  JHEP {\bf 0305} (2003) 007
  \texttt{[arXiv:hep-lat/0301020].}
\bibitem{nprpaper} Y.~Aoki et al. (RBC \& UKQCD Collaborations), paper in preparation.
\bibitem{braun}
  V.~M.~Braun, T.~Burch, C.~Gattringer, M.~Gockeler, G.~Lacagnina, S.~Schaefer and A.~Schafer,
\\ \textit{A lattice calculation of vector meson couplings to the
vector and tensor currents using chirally improved fermions},
  Phys.\ Rev.\  D {\bf 68} (2003) 054501
  [arXiv:hep-lat/0306006].
\bibitem{ukqcdqcdsf}
M.~Gockeler {\it et al.}, \textit{Meson decay constants from N(f)
= 2 clover fermions,}
  PoS {\bf LAT2005} (2006) 063
  [arXiv:hep-lat/0509196].
\bibitem{ms87} G.~Martinelli and C.~T.~Sachrajda,
\textit{A Lattice Calculation of the Pion's Form-Factor and
Structure Function,} Nucl.\ Phys.\  B {\bf 306} (1988) 865.
\bibitem{qcdsfgrouptheory} M.~Gockeler, R.~Horsley, E.~M.~Ilgenfritz, H.~Perlt, P.~Rakow, G.~Schierholz and A.~Schiller,
\textit{Lattice Operators for Moments of the Structure Functions
and their Transformation under the Hypercubic Group,}
  Phys.\ Rev.\  D {\bf 54} (1996) 5705
  [arXiv:hep-lat/9602029].
\bibitem{xiav2006} P.~A.~Boyle, M.~A.~Donnellan, J.~M.~Flynn, A.~Juttner,
J.~Noaki, C.~T.~Sachrajda and R.~J.~Tweedie, [UKQCD
Collaboration], \textit{A lattice computation of the first moment
of the kaon's distribution amplitude,}
  Phys.\ Lett.\  B {\bf 641} (2006) 67
  [arXiv:hep-lat/0607018].
\bibitem{chenstewart} J.~W.~Chen and I.~W.~Stewart,
\textit{Model independent results for SU(3) violation in
light-cone distribution functions,}
  Phys.\ Rev.\ Lett.\  {\bf 92} (2004) 202001
  [arXiv:hep-ph/0311285].
\bibitem{qcdsfpda} V.~M.~Braun {\it et al.},
\textit{Moments of pseudoscalar meson distribution amplitudes from
the lattice,}
  Phys.\ Rev.\  D {\bf 74} (2006) 074501
  [arXiv:hep-lat/0606012].



\end{thebibliography}
\end{document}